\documentclass[twocolumn]{aastex631}

\usepackage[normalem]{ulem}

\graphicspath{{./}{figures/}}

\def\deg{\hbox{$^\circ$}\,}

\def\FeIline{\ion{Fe}{1}~6173\,\AA\,}
\def\CaIR{\ion{Ca}{2}~8542\,\AA\,}

\def\Halpha{\mbox{H\hspace{0.1ex}$\alpha$}\,}
\def\logtau{$\log\tau_{\mathrm{500}}$\,}

\def\blos{$B_{\mathrm{LOS}}$\,}

\def\sv{Stokes~$V/I_{\mathrm{c}}$\,}
\def\si{Stokes~$I/I_{\mathrm{c}}$\,}
\def\deltab{$\Delta \lambda_{\mathrm{B}}$\,}
\def\deltad{$\Delta \lambda_{\mathrm{D}}$\,}

\begin{document}

\title{The \Halpha{} line as a probe of chromospheric magnetic fields}

\correspondingauthor{Harsh Mathur}
\email{harshmathur.1990@gmail.com}

\author[0000-0001-5253-4213]{Harsh Mathur}
\affiliation{Indian Institute of Astrophysics, II Block, Koramangala, Bengaluru 560 034, India}

\author[0000-0003-0585-7030]{Jayant Joshi}
\affiliation{Indian Institute of Astrophysics, II Block, Koramangala, Bengaluru 560 034, India}

\author[0000-0002-3208-6238]{Thore Espedal Moe}
\affiliation{Institute of Theoretical Astrophysics, University of Oslo, P.O. Box 1029 Blindern, NO-0315 Oslo, Norway}
\affiliation{Rosseland Centre for Solar Physics, University of Oslo, P.O. Box 1029 Blindern, NO-0315 Oslo, Norway}

\author[0000-0003-4747-4329]{Tiago M. D. Pereira}
\affiliation{Institute of Theoretical Astrophysics, University of Oslo, P.O. Box 1029 Blindern, NO-0315 Oslo, Norway}
\affiliation{Rosseland Centre for Solar Physics, University of Oslo, P.O. Box 1029 Blindern, NO-0315 Oslo, Norway}

\author[0000-0002-0465-8032]{K. Nagaraju}
\affiliation{Indian Institute of Astrophysics, II Block, Koramangala, Bengaluru 560 034, India}

\begin{abstract}
We explore the diagnostic potential of the \Halpha{} line for probing the chromospheric magnetic field using a realistic 3D radiative magnetohydrodynamic (rMHD) model.
The Stokes profiles of the \Halpha{} line are synthesized through full 3D radiative transfer under the field-free approximation, alongside the \CaIR{} and \FeIline{} lines for comparison.
The line-of-sight (LOS) magnetic fields are inferred using the weak field approximation (WFA) for the \Halpha{} and \CaIR{} lines, while the \FeIline{} line is analyzed through Milne-Eddington inversion techniques.
The comparison between the inferred LOS magnetic field maps and the magnetic fields in the rMHD model revealed that the \Halpha{} line core primarily probes the chromospheric magnetic field at \logtau{} = $-$5.7, which corresponds to higher layers than the \CaIR{} line core, which is most sensitive to conditions at \logtau{} = $-$5.1.
On average, the Stokes~$V$ profiles of the \Halpha{} line core form 500~km higher than those of the \CaIR{} line core.
The \Halpha{} polarization signals persist after adding noise, and with noise at the level of $10^{-3}I_c$, most simulated magnetic structures remain visible.
These findings suggest that spectropolarimetric observations of the \Halpha{} line can provide complementary insights into the stratification of the magnetic field at higher altitudes, especially when recorded simultaneously with widely used chromospheric diagnostics such as the \CaIR{} line.
\end{abstract}

\keywords{Solar magnetic fields, Solar chromosphere, Spectropolarimetry}

\section{Introduction}
\label{sect:introduction}

The \Halpha{} spectral line serves as a key diagnostic tool for probing the solar chromosphere and its dynamic phenomena, including filaments, Ellerman bombs, surges, flares, and spicules. 
These events exhibit distinct spectral signatures that manifest in the \Halpha{} intensity profiles, which are then used for the classification of solar features and for studying the fine structure and temporal evolution of the chromosphere.

Despite being a critical line to study the structure and evolution of the chromosphere, polarization studies to infer the magnetic field from this line remain limited.
For instance, \citet{1971SoPh...16..384A} utilized simultaneous measurements of the \ion{Fe}{1}~6302.5~\AA\, and \Halpha{} lines to estimate vertical magnetic field gradients.
Various studies have also documented simultaneous spectropolarimetric observations of the \Halpha{} line alongside \ion{Fe}{1} lines, enabling comparative analyses of photospheric and chromospheric magnetic fields in sunspots \citep[][]{2004ApJ...606.1233B, 2005PASJ...57..235H, 2008ApJ...678..531N}.
Radial variations of the line-of-sight magnetic field in both the chromosphere and photosphere of a sunspot were analyzed by \citet{2020JApA...41...10N}.
Furthermore, spectropolarimetric observations of the \Halpha{} line have been employed to diagnose magnetic fields in prominences \citep[][]{2005ApJ...621L.145L}.
Recently, \citet{2022A&A...659A.179J} explored linear and circular polarization signals near the North and South Solar Limb, inferring the line-of-sight magnetic field through the weak field approximation (WFA).

The reason the \Halpha{} line is not commonly used to infer chromospheric magnetic fields is due to its sensitivity to the complex 3D radiation field.
While the Zeeman effect largely influences the Stokes~$V$ signal, both the intensity and polarization profiles of this line are highly sensitive to the complex 3D radiation field \citep{2012ApJ...749..136L}, which makes accurate modeling—and consequently, the interpretation of its observations—highly challenging.
Additionally, in weakly magnetized atmospheres, the Stokes~$Q$~\&~$U$ signals are affected by atomic polarization \citep{2010MmSAI..81..810S, 2011ApJ...732...80S}, making the \Halpha{} line difficult to model with current inversion codes, which typically rely on a 1.5D plane-parallel geometry.
Furthermore, in weak-field conditions—where Zeeman splitting (\deltab{}) is significantly smaller than the Doppler width (\deltad{})—the amplitude of circular polarization is proportional to the ratio of \deltab{} to \deltad{}, while linear polarization scales with the square of this ratio \citep[see][p. 405]{2004ASSL..307.....L}).
This ratio is generally small for hydrogen, given its low atomic mass and correspondingly large Doppler width, in contrast to heavier atoms like calcium.

\citet{2004ApJ...603L.129S} previously calculated 1D response functions of the \Halpha{} line’s Stokes parameters, showing that it is notably sensitive to both photospheric and chromospheric magnetic fields.
However, \citet{2012ApJ...749..136L} demonstrated that a 1D radiative transfer approach is inadequate for accurately modeling the \Halpha{} line.
Only when the radiative transfer is treated in 3D, does the \Halpha{} line appear to trace chromospheric magnetic structures like fibrils.
Additionally, \citet[][]{2002ApJ...572..626C} showed that in the upper chromosphere, the \Halpha{} line’s opacity is mainly sensitive to the ionization degree and radiation field, which are largely insensitive to the local temperature variations but are sensitive to mass density  \citep[][]{2012ApJ...749..136L}.
Recent work by \citet{2019A&A...631A..33B} further confirmed the persistence of \Halpha{} opacity in flaring active regions by synthesizing spectra using 3D radiative MHD simulations.
Consequently, simultaneous spectropolarimetric observations of the \Halpha{} line with well-established chromospheric diagnostics, such as those of the \ion{Ca}{2} and \ion{He}{1} atoms \citep[][]{2017SSRv..210...37L}, present a robust probe for investigating chromospheric magnetic fields.

\begin{figure}[htbp]
    \centering
    \includegraphics[width=0.5\textwidth]{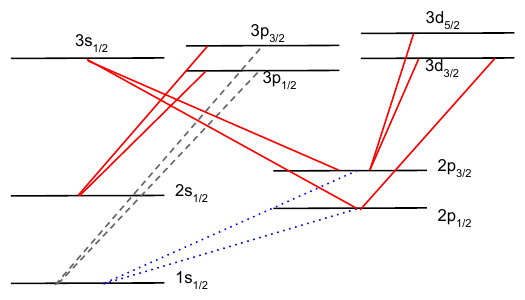}
    \caption{Grotrian diagram of Hydrogen atom showing the \Halpha{} line is a combination of 7 transitions. The solid red-colored transitions correspond to the \Halpha{} line, while the dotted blue-colored transitions correspond to Ly $\alpha$ and the dashed grey-colored transitions correspond to the Ly $\beta$ line.}
    \label{fig:grotrian}
\end{figure}

Recent studies by \citet{2023ApJ...946...38M, 2024ApJ...971...30M} have utilized simultaneous spectropolarimetric observations of the \Halpha{} and \ion{Ca}{2}~8542~\AA\, (and 8662~\AA) lines to investigate the stratification of the chromospheric magnetic field by applying the weak-field approximation to the line core and wings, separately.
In the study, \citet{2023ApJ...946...38M} found that the line-of-sight (LOS) magnetic field morphology in the \Halpha{} line core closely resembles that of the \CaIR{} line, though it appeared more diffuse, and the inferred magnetic field strength was consistently weaker than that of the \CaIR{} line.
However, these observations were limited to a small pore.
In the subsequent study, of a complex active region, \citet{2024ApJ...971...30M} observed that the LOS magnetic fields inferred from the \Halpha{} line core were uncorrelated, similarly diffuse, and consistently weaker than those from the \ion{Ca}{2}~8662~\AA\, and \ion{Fe}{1}~8661.8991~\AA\, lines.
The weaker and more diffuse fields inferred from the line core, combined with the assumption that magnetic fields expand with height, led the authors to conclude that the \Halpha{} line core is a reliable probe of the chromospheric magnetic field.

In this letter, we present a theoretical investigation into the diagnostic potential of the \Halpha{} line for probing the chromospheric magnetic field.
Using a publicly available 3D radiative magnetohydrodynamic (rMHD) simulation, we synthesize the Stokes profiles of the \Halpha{} line through full 3D radiative transfer in field-free approximation.
For comparison, we also synthesize the Stokes profiles of the \CaIR{} and \FeIline{} lines.
To infer the LOS magnetic field, we apply the WFA to the \Halpha{} and \CaIR{} lines, and the Milne-Eddington inversion to the \FeIline{} line.
We then compare the morphology and field strengths of the inferred LOS magnetic field maps with the true fields in the 3D rMHD model.

\section{Modeling the polarization profiles of the \Halpha{}, \CaIR{} and \FeIline{} lines}
\label{sect:halphamodel}

\begin{figure*}[htbp]
    \centering
    \includegraphics[width=\textwidth]{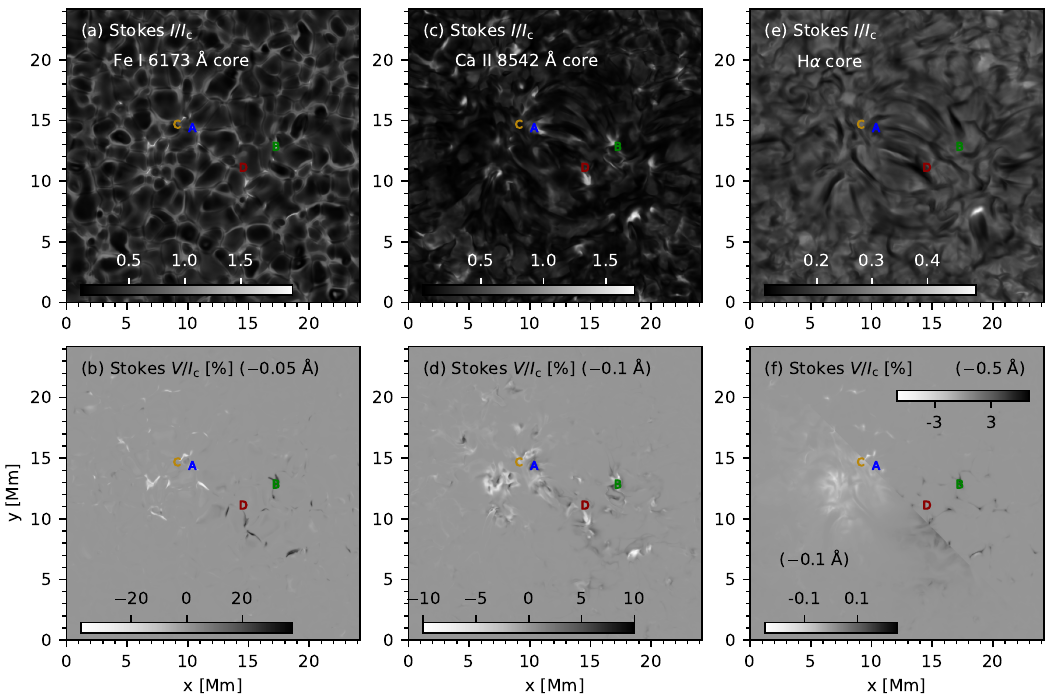}
    \caption{
    Synthetic \si{} and \sv{} maps. The \si{} maps in the top row (panels (a),(c), and (e)) are shown at the line core of the \FeIline{}, \CaIR{}, and \Halpha{} lines. Panels (b) and (d) represent \sv{} at $-$0.5~\AA\ and $-$0.1~\AA\ for the \FeIline{} and \CaIR{} line, respectively.  The top-right diagonal half of panel (f) displays the \sv{} image at $-$0.5~\AA\ from the \Halpha{} line core, while the bottom-left diagonal half presents the \sv{} image at $-$0.1~\AA\ from the \Halpha{} line core. Spectral profiles at a few selected pixels are shown in Fig.~\ref{fig:profcontext}.}
    
    \label{fig:context}
\end{figure*}

To model the polarization profiles of the \Halpha{} line in 3D, we have used a publically available enhanced network simulation \citep{2016A&A...585A...4C} computed using the Bifrost Code \citep{2011A&A...531A.154G}.
This simulation used an equation of state that includes the effects of non-equilibrium ionization of Hydrogen.
Using the simulation, we computed the hydrogen level populations using the \textit{Multi3d code} \citep{2009ASPC..415...87L}, assuming a 6-level hydrogen atom (5 bound $n$-levels plus continuum), the same as those used in the \citet{2012ApJ...749..136L} study.
While the version of \textit{Multi3d} we used can compute intensity profiles in 3D, it does not handle polarization.
Additionally, to compute polarization, fine-structures ($nlj$) states of $n$ levels must be included in the computation.
The \Halpha{} line is a combination of 7 transitions between the fine-structure states of $n=3$ to $n=2$ as shown in Fig.~\ref{fig:grotrian}.
To account for this, we redistributed the populations into a 9-level hydrogen atom, as illustrated in Fig.~\ref{fig:grotrian}.
The population of each $n$-level was redistributed across the fine-structure $nlj$-states in proportion to their statistical weights, ensuring that the density matrix elements of all $nlj$-states corresponding to the same $n$-level remain the same. 
These populations are then held constant for the subsequent calculation of polarization profiles, an approach known as the ``field-free'' approximation.
%
%
We used the collisional rates given by \citet{2004ApJ...609.1181P} for this model hydrogen atom. 
This model atom is then used with the RH code \citep{2001ApJ...557..389U}, on a column-by-column basis (also known as 1.5D approximation) to do just the formal solution to calculate the emergent intensity and polarization profiles along the line-of-sight (LOS) at $\mu$ = 1.
We reiterate that the above step done using the RH code, did not modify the level populations.

To calculate the polarization profiles of the \FeIline{} and \CaIR{} lines, the RH code is again used under the 1.5D approximation.
The \FeIline{} is synthesized using Kurucz data \citep{2011CaJPh..89..417K} under the local thermodynamic equilibrium (LTE) approximation.
Recent studies have inferred the LOS magnetic field from the \FeIline{} line using the LTE approximation \citep[for eg.][]{2020A&A...635A.210P, 2021A&A...649A.106Y, 2022A&A...664A..72J, 2022A&A...668A.153M}.
However, we note that \citet{2023A&A...669A.144S}, using a realistic 3D MHD simulation of a non-grey sunspot cube \citep{2012ApJ...750...62R} generated with the MURaM code \citep{2005A&A...429..335V}, showed that LTE approximation may lead to underestimating the LOS magnetic field.
The \ion{Ca}{2}~IR lines were synthesized with the complete re-distribution (CRD) approximation using a 6-level \ion{Ca}{2} atom.
The polarization profiles of the \CaIR{} line were also synthesized in ``field-free'' approximation, because we do not expect the magnetic field to influence the populations.
While solving the radiative transfer for the \CaIR{} and \FeIline{} lines, the populations of the $nlj$-levels of the hydrogen atom were kept fixed to the populations calculated from the \textit{Multi3d} code, as explained in the previous paragraph.

\section{Description of synthetic Stokes~$I$ and $V$ maps and profiles of the \Halpha{} line}\label{sect:descsp}

\begin{figure*}[htbp]
    \centering
    \includegraphics[width=\textwidth]{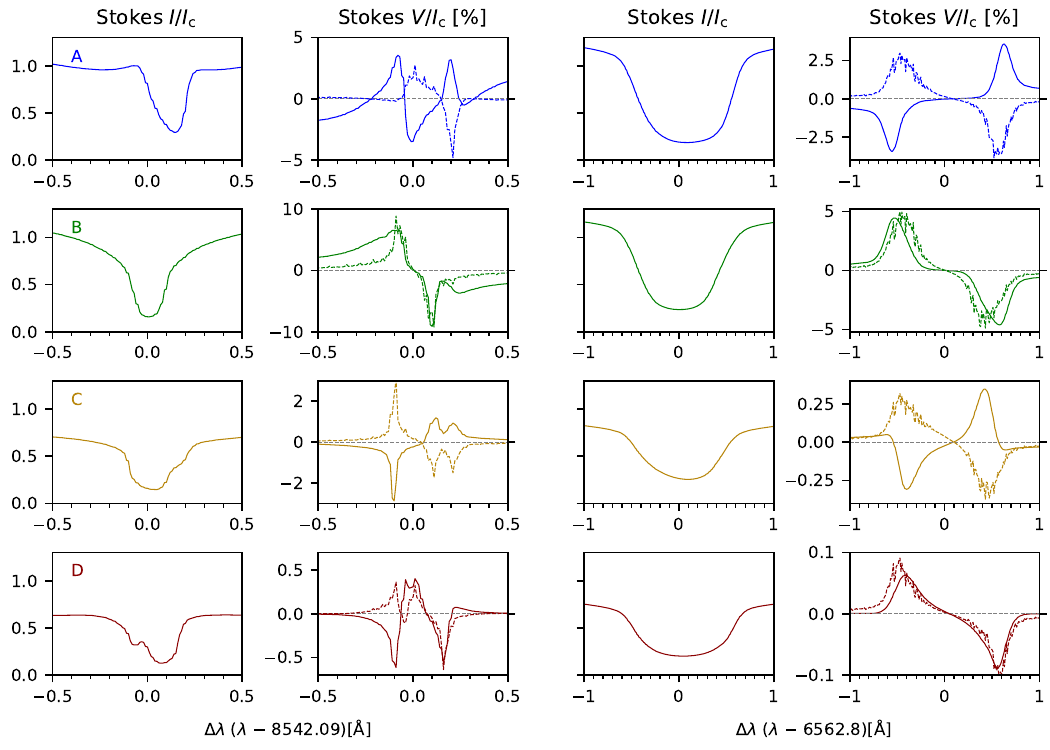}
    \caption{\si{} and $V/I_{\mathrm{c}}$ profiles of a few selected pixels. The amplitudes of \si{} and \sv{} are marked on ticks on the left-axis of each of the subplots. The dashed lines show the derivative of the \si{} with the sign changed, to compare with the \sv{}. The scale of the derivatives is not shown.}
    \label{fig:profcontext}
\end{figure*}

Figure~\ref{fig:context} shows the synthesized intensity and Stokes~$V/I_{\mathrm{c}}$ maps in the \FeIline{}, \CaIR{} and \Halpha{} lines.
Where  $I_{\mathrm{c}}$ represents the median intensity calculated separately for each spectral line across a region in the FOV with no magnetic activity, at a wavelength point $-$4~\AA\, away from the line center for the \CaIR{} and \Halpha{} lines, and $-$2~\AA\, away from the line center for the \FeIline{} line.
The \si{} image for the \FeIline{} line core displays a reverse granulation pattern, while the Stokes $V/I_{\mathrm{c}}$ image reveals the structure of the magnetic field network and magnetic field concentrated in intergranular lanes (see panels a and b), with a maximum amplitude of about 36\%.
For the \CaIR{} line core, the \si{} image captures the chromospheric fibrilar structure, whereas the \sv{} image depicts a diffuse magnetic field structure indicative of the chromosphere (see panels c and d), with a maximum amplitude of approximately 9.5\%.
In the \Halpha{} line core, the \si{} image resembles the top-left panel of Fig.~7 from \citet{2012ApJ...749..136L}.
The \sv{} map of the \Halpha{} line core at the wavelength position corresponding to the maximum amplitude of the \sv{} profiles (see Fig.\ref{fig:profcontext}), $-$0.5~\AA, is shown in the top-right diagonal half of panel (f).
This map exhibits a pattern similar to the network fields found in the simulation, with a maximum amplitude of 6.5\%.
%
The map at $-$0.1~\AA\, bottom-left diagonal half of panel (f), shows a more diffuse structure than that in the Stokes $V/I_{\mathrm{c}}$ map (panel (d)) for the \CaIR{} line. The $V/I_{\mathrm{c}}$ map at $-$0.1~\AA\, in the \Halpha{} line has maximum amplitude of 0.2\%.
To compare with active region observations reported in \citet{2024ApJ...971...30M}, the maximum amplitude observed in the \sv{} profiles of the \Halpha{} line at the wavelength position corresponding to the maximum amplitude of Stokes~$V$ was observed to be 3\%.

The spectral profiles of 4 selected pixels, marked by `A', `B', `C' and `D' in Fig.~\ref{fig:context}, are shown in Fig.~\ref{fig:profcontext}.
The \si{} profiles for the \CaIR{} and \Halpha{} lines are plotted in the first and third columns, respectively, while the \sv{} profiles for these lines are presented in the second and fourth columns.
Additionally, the derivative of the \si{} ($\frac{d I/I_{\mathrm{c}}}{d \lambda}$, shown with dashed lines) is overplotted on the \sv{} profiles. 
The stratification of the LOS magnetic field of these 4 pixels is shown in panels (A), (B), (C), and (D) of Fig.~\ref{fig:response} of the Appendix~\ref{sect:appresp}, respectively.
In this simulation, the LOS magnetic field follows the convention where positive values indicate an outward-directed field, consistent with the positive direction of the $z$-axis.
%

The spectral profiles shown in blue color, pixel `A', represent a pixel where the LOS magnetic field consistently has a negative sign throughout all heights of the atmosphere.
%
%
Notably, the \sv{} profile of the \CaIR{} line exhibits multiple lobes and sign changes from the wings to the core, which might give an impression that magnetic field polarity is reversed from the photosphere to the chromosphere.
However, this sign change aligns with the sign change in the derivative of \si{} of the \CaIR{} line, indicating no change in magnetic polarity with height.
%
%
The \si{} profiles of the \Halpha{} line display nominal absorption, and the blue wing of the \sv{} profile for the \Halpha{} line shows a negative sign, accurately indicating the pixel's negative magnetic field polarity.

The spectral profiles shown in green color, pixel `B', represent a pixel where the LOS magnetic field consistently has a positive sign.
The \si{} profiles of both the \CaIR{} and \Halpha{} lines show nominal absorption and blue wing of the \sv{} profiles show a positive sign, accurately indicating the pixel's positive magnetic polarity.

The spectral profiles shown in brown color, pixel `C', represent a pixel where the LOS magnetic field is positive in the lower photospheric layers (\logtau{} $>$ $-$1.8) and negative in the upper photospheric and chromospheric layers (\logtau{} $<$ $-$1.8).
%
The corresponding \CaIR{} \sv{} profile is complex showing multiple lobes. 
The blue lobe of the \sv{} profile for the \CaIR{} line shows only a negative sign, which does not reflect the positive polarity of the lower photosphere. 
However, it is possible that a signature of the lower photospheric magnetic field might be hidden in this complex profile.
Or it may also be possible that the wings of the \CaIR{} line probe higher photospheric heights than those probed by the \FeIline{} line.
%
%
In contrast, the \Halpha{} line has a typical two lobed \sv{} profile. 
Moreover, the sign of \sv{} profiles shifts from positive to negative around $\sim$$-$0.5~\AA, clearly indicating that the \Halpha{} line is capturing magnetic polarity reversal with height with wings sampling the photospheric fields, while the line core samples the chromospheric fields.

The spectral profiles shown in red, pixel `D', correspond to a pixel where the LOS magnetic field exhibits a negative sign in the lower chromospheric layers (\logtau{} $\sim$ $-$5) and a positive sign in the upper chromospheric layers (\logtau{} $\sim$ $-$5.7).
The blue lobe of the \sv{} profile of the \CaIR{} line displays a negative sign, indicating that it samples the lower chromospheric layers.
Interestingly, the blue lobe of the \sv{} profile of the \Halpha{} line shows a positive sign, suggesting that the \Halpha{} line may probe magnetic fields at higher chromospheric heights than those sampled by the \CaIR{} line.

\section{Methods of analysis}
\label{sect:methods}

\subsection{Weak field approximation}

The apparent LOS magnetic field of the \CaIR{} line and \Halpha{} was inferred using the weak field approximation method (WFA).
We have used a spatially-coupled version of WFA developed by \citet{2020A&A...642A.210M}, which imposes spatial coherency in the WFA results.
The magnetic field for the \CaIR{} line core and the \Halpha{} line core was calculated within the wavelength range of $\pm$0.25~\AA\, and $\pm$0.15~\AA, respectively.
Using a narrower wavelength range (closer to line core) for the \CaIR{} line does not result in sensitivity to magnetic fields higher up in the atmosphere, but it makes the measurement more noisy, so we opted to use $\pm$0.25~\AA\, for this line.
As discussed in section~\ref{sect:descsp}, the Stokes~V map at the wavelength position where the amplitude of the Stokes~$V$ profile of the \Halpha{} line is maximum, $-$0.5~\AA, resembles network fields, thus shows a significant photospheric contribution.
Thus, we have chosen $\pm$0.15~\AA\, as the wavelength range representing the core of the \Halpha{} line.
Moreover, choosing a broader wavelength range for the WFA of the \Halpha{} line results in photospheric contribution visible in the magnetic field maps.
Whereas, the magnetic field for the wings of the \CaIR{} line and the \Halpha{} line was determined over the wavelength ranges of [$-$2, $-$1]~\AA\, and [$+$0.35, $+$1.5]~\AA, respectively.
More details about applying the WFA on the \CaIR{} and \Halpha{} lines can be found in \citet{2023ApJ...946...38M, 2024ApJ...971...30M}. 

\subsection{Milne-Eddington Inversion}

We performed Milne-Eddington (ME) inversions \citep[see chapter 11 of][]{2007insp.book.....D} of the \FeIline{} Stokes profiles to infer the \blos{} utilizing pyMilne code, a parallel C++/Python implementation\footnote{\url{https://github.com/jaimedelacruz/pyMilne}} \citep{2019A&A...631A.153D}.

\subsection{Response functions} \label{sect:resp}

\begin{figure*}[htbp]
    \centering
    \includegraphics[width=\textwidth]{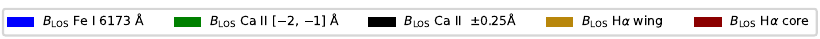}
    \includegraphics[width=\textwidth]{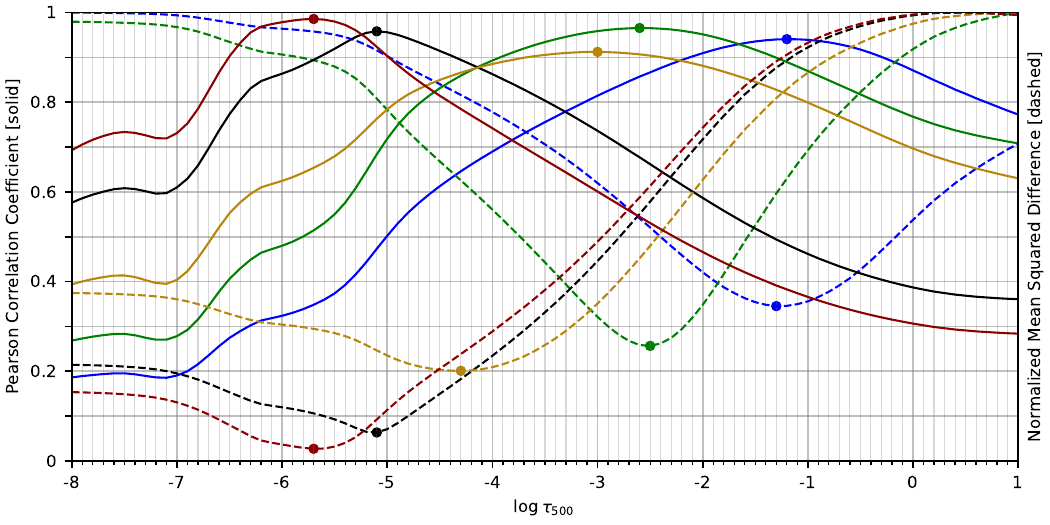}
    \caption{Pearson correlation (solids) and normalized mean squared difference (dashed) of the inferred magnetic fields with the 3D MHD model at different \logtau{} values.}
    \label{fig:tauformation}
\end{figure*}


We calculated the response functions to the \blos{} by perturbing the LOS magnetic field at various heights and synthesized the profiles for those perturbed atmospheres.
To calculate response functions, ``$+$'' and ``$-$'' perturbations ($\Delta x$) are applied to the \blos{} at each height ($h$), resulting in the synthesized spectral line Stokes profiles, $S^{+}$ and $S^{-}$, respectively.
The response of the \blos{} at each height is then determined by
\begin{equation}
RF_{B_{\mathrm{LOS}}}(h, \lambda) = \frac{S^{+} - S^{-}}{2 * \Delta x}.
\end{equation}

%
We do not expect the magnetic field to influence the level populations; thus, for the \Halpha{} line, we do not perform any iterations in the RH code, as explained in section~\ref{sect:halphamodel}.
%
Instead, we directly apply the formal solution to compute the emergent intensity and polarization profiles for the perturbed atmospheres.
For the \CaIR{} and \FeIline{} lines, we re-ran the RH 1.5D code for each of the perturbed atmospheres, because it was easier that way.

\section{Results and discussions}
\label{sect:results}

\subsection{Mapping atmospheric layers: magnetic field sensitivity of the \Halpha{} line} \label{sect:mapping}

One of the primary objectives of this study is to determine the atmospheric height at which the \Halpha{} line exhibits sensitivity to the magnetic field.
Additionally, we explore the \Halpha{} line as a probe of the magnetic field and compare its response with the \CaIR{} and \FeIline{} lines to assess their diagnostic potential across different atmospheric layers. 
We compared the LOS magnetic field maps retrieved from all the three lines deploying various methods described in the section~\ref{sect:methods} to the actual magnetic field at different optical depth in the simulation.
%
%
Figure~\ref{fig:tauformation} represents the Pearson correlation coefficients between the inferred LOS magnetic fields maps and the LOS magnetic field maps at different \logtau{} values in the 3D MHD simulation. 
Additionally, the normalized mean squared difference between the inferred LOS magnetic fields and the 3D MHD model at different \logtau{} values is also shown. 
A higher Pearson correlation coefficient indicates a stronger spatial correlation between the inferred magnetic field maps and those from the 3D MHD model.
Whereas, lower values of mean squared difference reflect a closer match between the inferred magnetic field amplitude and the model.

\begin{figure*}[htbp]
    \centering
    \includegraphics[width=0.9\textwidth]{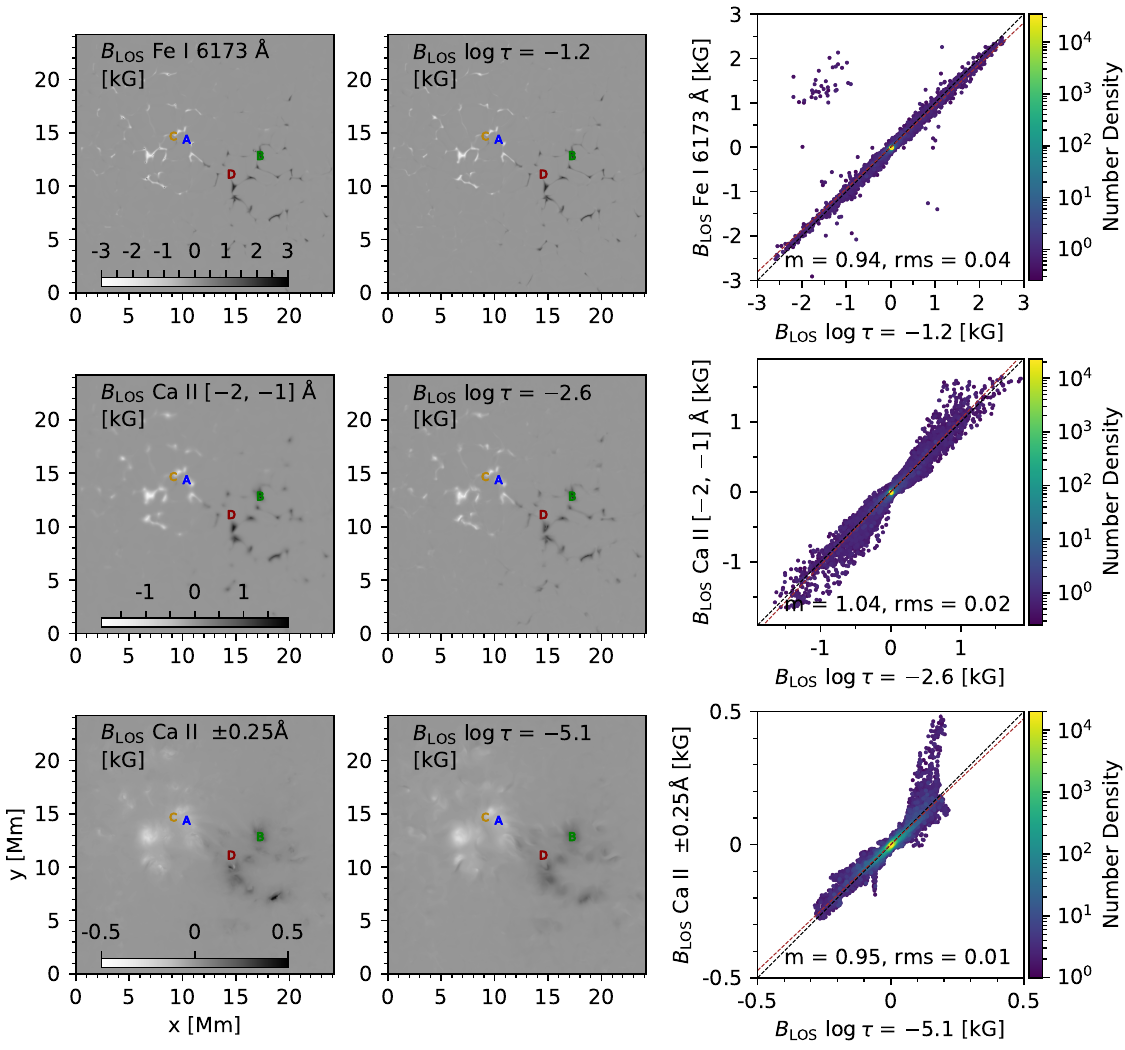}
    \caption{Comparison of magnetic fields inferred from \FeIline{} and \CaIR{} lines with those of the 3D MHD model. The first column shows magnetic field maps derived from synthetic observables. The second column displays LOS magnetic field maps from the 3D MHD model at the atmospheric layer at which the corresponding maps in the first column show the best correlation (see Fig.~\ref{fig:tauformation}). The final column presents a 2D histogram between the maps in the first and second columns. The slope of the linear regression of the scatter plot in the third column is indicated by $m$ and rms indicates the root mean square of the difference between the two values.}
    \label{fig:fecacomparisonmap}
\end{figure*}

The blue-colored curve represents the LOS magnetic field derived from the \FeIline{} line, showing the highest spatial correlation with the 3D MHD model at \logtau{} = $-$1.2 and the lowest mean squared difference at \logtau{} = $-$1.3.
This aligns with the well-established understanding that \ion{Fe}{1} lines typically sample the photospheric magnetic field \citep{1982A&A...115..104R, 1988ASSL..138..185R}.
The green-colored curve corresponds to the magnetic field map inferred using WFA from the wings of the \CaIR{} line, with maximum Pearson correlation at \logtau{} = $-$2.6 and the lowest mean squared difference at \logtau{} = $-$2.5.
In earlier studies using semi-empirical models \citep[FALC;][]{1985cdm..proc...67A, 1993ApJ...406..319F}, \citet{2018ApJ...866...89C} identified that the wings of the \CaIR{} line sample photospheric layers around \logtau{} $\sim$ $-$1.4.
This discrepancy may arise because \citet{2018ApJ...866...89C} employed a semi-empirical model that does not inherently include a magnetic field and applied different magnetic field stratifications in their analysis, whereas our study utilizes a realistic 3D rMHD model.

\begin{figure*}[htbp]
    \centering
    \includegraphics[width=0.9\textwidth]{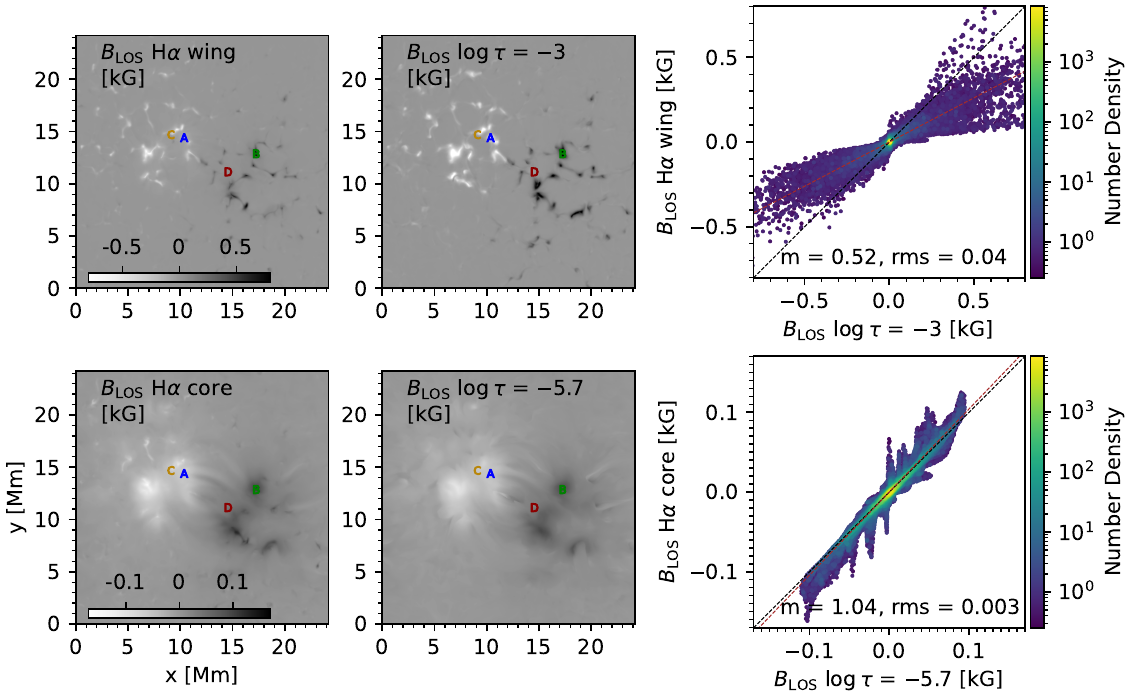}
    \caption{Comparison of magnetic fields inferred from \Halpha{} line with that of the 3D MHD model in the same format as in Fig.~\ref{fig:fecacomparisonmap}.}
    \label{fig:hacomparisonmap}
\end{figure*}

The black-colored curve reflects the maps derived from the core of the \CaIR{} line, having the highest spatial correlation and minimum mean squared difference at \logtau{} = $-$5.1.
According to \citet{2018ApJ...866...89C}, the core of the \CaIR{} line is sensitive around \logtau{} = $-$5.3.

In contrast to the \FeIline{} and \CaIR{} lines, the magnetic field map retrieved from the wings of the \Halpha{} line (brown curve) shows spatial correlation and the lowest mean squared difference at notably different \logtau{} values, specifically $-$3 and $-$4.2, respectively.
The higher spatial correlation at \logtau{} = $-$3 suggests influence from upper photospheric or lower chromospheric layers, while the minimum mean squared difference at \logtau{} = $-$4.2 indicates some contribution from even higher layers. 
Therefore it suggests that magnetic field inferred form the \Halpha{} wings might sample a range of the atmosphere making its interpretation more complex.
The core of the \Halpha{} line shows the highest spatial correlation and lowest mean squared difference both at \logtau{} = $-$5.7, suggesting that the LOS magnetic field maps inferred from the \Halpha{} line core consistently probe chromospheric magnetic fields at greater heights than those probed by the \CaIR{} line core (\logtau{} = $-$5.1).

With the above analysis, we conclude that \Halpha{} line core and wings sample the LOS magnetic field around \logtau{} = $-$5.7 and \logtau{} = $-$3.0, respectively. 
In comparison, the \CaIR line core and wings are more sensitive to the LOS magnetic field around \logtau{} = $-$5.1 and \logtau{} = $-$2.6, respectively. 
The LOS magnetic field inferred using the Milne-Eddington inversion of the \FeIline{} line represents the magnetic field at  \logtau{} = $-$1.2. 
In Fig.~\ref{fig:fecacomparisonmap} and \ref{fig:hacomparisonmap}, we compare the LOS magnetic field maps inferred from the synthetic observables with that of the ground truth, the 3D MHD model.
%
%
%
%
The morphology of the LOS magnetic field derived from \FeIline{}, as well as from the wings and core of the \CaIR{} line, aligns closely with the LOS magnetic field maps at layers where the Pearson correlation is highest.
The scatter plots in the third column indicate that the inferred amplitudes correspond well with the model, with slopes close to unity.
The rms differences between the inferred maps and the 3D MHD model are 40~G for \FeIline{}, 20 G for the wings of \CaIR{}, and 10~G for the core of \CaIR{}.

The LOS magnetic field inferred from the wings of the \Halpha{} line shows a weak correlation with the layer at which the Pearson correlation is highest, especially when compared to the maps derived from the \FeIline{} and \CaIR{} lines, and exhibits lower amplitudes than the 3D MHD model.
This finding is also reflected in the scatter plot in the third column, which displays a significant scatter and a slope of around 0.5.
However, the rms differences, approximately 40~G, are comparable to those in the maps inferred from the \FeIline{} and \CaIR{} lines.
The correlation with the lower chromospheric / upper photospheric layers, \logtau{} = $-$3, though with lower field amplitudes, suggests the contribution of photospheric fields in addition to chromospheric fields within the Stokes~$V$ profiles of the wings of the \Halpha{} line.

The LOS magnetic field maps inferred from the core of the \Halpha{} line, unlike those from the wings, exhibit a strong spatial correlation and similar amplitudes to the \logtau{} = $-$5.7 layer of the 3D MHD model.
This is also evident in the scatter plot in the third column, which shows a unity slope and rms differences of just 3~G.
This result strongly supports the hypothesis that the core of the \Halpha{} line consistently probes the chromospheric magnetic field at higher layers than the \CaIR{} line, which probes \logtau{} $\sim$ $-$5.

To assess the effect of noise in high-resolution ground-based solar observations, we simulate realistic noise conditions and re-infer magnetic fields using the weak-field approximation.
While the \Halpha{} maps appear noisier due to weaker polarization signals, the magnetic structures remain discernible, with an rms error of 18~G, and the scatter decreases with increasing field strength—indicating strong potential for application in active regions.
More details are provided in appendix~\ref{sect:dkist}.

\subsection{Analysis of response functions and formation height of polarization profiles of the \Halpha{} line}

To determine the optical depths (heights) where the Stokes~$V$ profile of the \Halpha{} line is most sensitive to the LOS magnetic field, we analyzed the response functions of the \Halpha{}, \CaIR{}, and \FeIline{} lines for a few representative pixels marked in Fig.~\ref{fig:context}.
The maximum response of these lines occurs at different \logtau{} values: \Halpha{} line core at $-5.9$ to $-$5.6, \CaIR{} line at $-5.7$ to $-$5.1, and \FeIline{} line at $-1.2$.
For pixels with simple magnetic field stratification (same polarity across heights), the inferred LOS magnetic field closely matches the rMHD model at the \logtau{} with maximum response.
However, for pixels with complex stratification (steep gradients and polarity reversals), the inferred field remains close to the rMHD model but may differ from the value of rMHD model at the \logtau{} with maximum response.
This shift is attributed to the complex Stokes~$I$ and $V$ profiles observed in these regions.
The temperature structure and the velocity fields may have contributed to the formation of such profiles in addition to the stratification of the LOS magnetic fields.
Overall, the analysis shows that the Stokes~$V$ profiles of the \Halpha{} line core are sensitive to magnetic field perturbations at higher atmospheric layers than those of the \CaIR{} line.
Consequently, the \Halpha{} line core probes the LOS magnetic field at higher layers than that inferred from the \CaIR{} line.
A detailed analysis, including the response functions and magnetic field stratifications for individual pixels, is provided in Appendix~\ref{sect:appresp}.

\begin{figure}[htbp]
    \centering
    \includegraphics[width=0.5\textwidth]{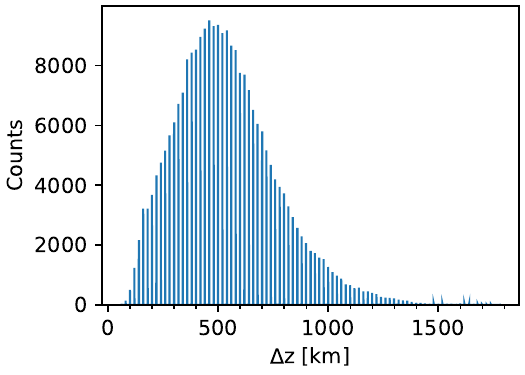}
    \caption{Histogram showing the difference of height of formation between the \Halpha{} and \ion{Ca}{2} line.}
    \label{fig:heighthist}
\end{figure}


In Fig.~\ref{fig:heighthist}, we show an histogram of difference between the height at \logtau{} = $-$5.7 with the height at \logtau{} = $-$5.1.
This difference can be as minimal as 100~km and can be as high as 1800~km.
On average, this difference is about 500~km.
Thus, it clearly shows that the Stokes~$V$ profiles of the \Halpha{} line core form on average 500~km above that of the \CaIR{} line.

\section{Summary and conclusions}
\label{sect:conclusions}

We have investigated whether the Stokes~$I$ and $V$ profiles of the \Halpha{} line probe the chromospheric magnetic field.
We modeled the Stokes~$I$ and $V$ profiles of the \Halpha{} line using 3D radiative transfer using the field-free approximation and compared the LOS magnetic field inferred with the 3D rMHD model used.
In addition to the \Halpha{} line, we have also synthesized the Stokes~$I$ and $V$ profiles of the \CaIR{} and \FeIline{} lines, for comparison.

The Pearson correlation coefficient between the LOS magnetic field map inferred from the \Halpha{} line core and the rMHD model is highest at \logtau{} = $-$5.7, whereas for the \CaIR{} line core, the highest correlation occurs at \logtau{} = $-$5.1.
This is clear evidence in support of conclusions from earlier observational studies, \citet{2023ApJ...946...38M} and \citet{2024ApJ...971...30M}, that the \Halpha{} line core consistently probes chromospheric magnetic field at higher layers than the \CaIR{} line.

The changes in the polarity of the LOS magnetic field across different layers of the solar atmosphere are reflected on the Stokes~$V$ profiles of the \Halpha{} and \CaIR{} lines.
When the magnetic field changes sign from the photosphere to the chromosphere, we observe that the sign of the Stokes~$V$ also changes from the line wings to the line core.
Similarly and interestingly, when there was a change in polarity at different heights of the solar chromosphere, the blue lobe of the Stokes~$V$ profiles of the \Halpha{} and \CaIR{} line core had different signs.

In contrast to the \Halpha{} line core, the map of the LOS magnetic field inferred from the wings of the \Halpha{} line had higher spatial correlation at \logtau{} = $-$3, whereas the minimum mean squared difference was at  \logtau{} = $-$4.2.
This suggests that the \Halpha{} line wings have significant contributions from both photosphere and chromospheric heights, making it challenging to interpret.

The study of the response functions also confirms that the maximum sensitivity of the Stokes~$V$ profiles of the \Halpha{} line core to the perturbations of the LOS magnetic field is at higher atmospheric layers (\logtau{} $\in$ [$-5.9$, $-$5.6]) than the \CaIR{} line (\logtau{} $\in$ [$-5.7$, $-$5.1]).
On average, the Stokes~$V$ profiles of the \Halpha{} line core forms about $+$500~km above the Stokes~$V$ profiles of the \CaIR{} line, ranging from $+$100 to $+$1800~km.
Adding noise at the level of $10^{-3}I_c$ produces noisier \Halpha{} polarization maps, but we find that most simulated magnetic structures remain visible.
The method is robust even for weaker fields.

In summary, the Stokes~$V$ profiles of the \Halpha{} line core always probe the chromospheric magnetic field at higher layers than the \CaIR{} line.
Spectropolarimetric observations of the \Halpha{} line when recorded simultaneously with widely used chromospheric diagnostics such as the \CaIR{} line can provide complementary insights into the stratification of the magnetic field by probing LOS magnetic field at higher layers than the \CaIR{} line.
Instruments like the Visible Spectro-Polarimeter \citep{2022SoPh..297...22D} on the Daniel K. Inouye Solar Telescope \citep{2020SoPh..295..172R} offer the capability to conduct such simultaneous observations.

\begin{acknowledgements}
This research has made use of the High-Performance Computing (HPC) resources (NOVA cluster) made available by the Computer Center of the Indian Institute of Astrophysics, Bangalore.
J.J. acknowledges financial support from the Science and Engineering Research Board (SERB)/Anusandhan National Research Foundation (ANRF), India, under the Core Research Grant (CRG), Grant No. CRG/2023/007464.
%
This work has been supported by the Research Council of Norway through its Centers of Excellence scheme, project number 262622.
Computational resources have been provided by UNINETT Sigma2 - the National Infrastructure for High Performance Computing and Data Storage in Norway.
This research has made use of NASA’s Astrophysics Data System Bibliographic Services.
We have also used the packages h5py \citep{collette_python_hdf5_2014}, matplotlib \citep{Hunter2007}, and NumPy \citep{harris2020array} to carry out our data analysis.

\end{acknowledgements}

\software{NumPy\citep{harris2020array},
          matplotlib\citep{Hunter2007},
          RH\citep{2001ApJ...557..389U},
          \textit{Multi3d}\citep{2009ASPC..415...87L},
          h5py \citep{collette_python_hdf5_2014},
          pyMilne\citep{2019A&A...631A.153D},
          SpatialWFA\citep{2020A&A...642A.210M}
          }

\bibliography{sample631}{}

\begin{thebibliography}{}
\expandafter\ifx\csname natexlab\endcsname\relax\def\natexlab#1{#1}\fi
\providecommand{\url}[1]{\href{#1}{#1}}
\providecommand{\dodoi}[1]{doi:~\href{http://doi.org/#1}{\nolinkurl{#1}}}
\providecommand{\doeprint}[1]{\href{http://ascl.net/#1}{\nolinkurl{http://ascl.net/#1}}}
\providecommand{\doarXiv}[1]{\href{https://arxiv.org/abs/#1}{\nolinkurl{https://arxiv.org/abs/#1}}}

\bibitem[{{Abdussamatov}(1971)}]{1971SoPh...16..384A}
{Abdussamatov}, H.~I. 1971, \solphys, 16, 384, \dodoi{10.1007/BF00162480}

\bibitem[{{Avrett}(1985)}]{1985cdm..proc...67A}
{Avrett}, E.~H. 1985, in Chromospheric Diagnostics and Modelling, ed. B.~W. {Lites}, 67--127

\bibitem[{{Balasubramaniam} {et~al.}(2004){Balasubramaniam}, {Christopoulou}, \& {Uitenbroek}}]{2004ApJ...606.1233B}
{Balasubramaniam}, K.~S., {Christopoulou}, E.~B., \& {Uitenbroek}, H. 2004, \apj, 606, 1233, \dodoi{10.1086/383118}

\bibitem[{{Bj{\o}rgen} {et~al.}(2019){Bj{\o}rgen}, {Leenaarts}, {Rempel}, {Cheung}, {Danilovic}, {de la Cruz Rodr{\'\i}guez}, \& {Sukhorukov}}]{2019A&A...631A..33B}
{Bj{\o}rgen}, J.~P., {Leenaarts}, J., {Rempel}, M., {et~al.} 2019, A\&A, 631, A33, \dodoi{10.1051/0004-6361/201834919}

\bibitem[{{Carlsson} {et~al.}(2016){Carlsson}, {Hansteen}, {Gudiksen}, {Leenaarts}, \& {De Pontieu}}]{2016A&A...585A...4C}
{Carlsson}, M., {Hansteen}, V.~H., {Gudiksen}, B.~V., {Leenaarts}, J., \& {De Pontieu}, B. 2016, \aap, 585, A4, \dodoi{10.1051/0004-6361/201527226}

\bibitem[{{Carlsson} \& {Stein}(2002)}]{2002ApJ...572..626C}
{Carlsson}, M., \& {Stein}, R.~F. 2002, \apj, 572, 626, \dodoi{10.1086/340293}

\bibitem[{{Centeno}(2018)}]{2018ApJ...866...89C}
{Centeno}, R. 2018, \apj, 866, 89, \dodoi{10.3847/1538-4357/aae087}

\bibitem[{Collette(2013)}]{collette_python_hdf5_2014}
Collette, A. 2013, Python and HDF5 (O'Reilly)

\bibitem[{{de la Cruz Rodr{\'\i}guez}(2019)}]{2019A&A...631A.153D}
{de la Cruz Rodr{\'\i}guez}, J. 2019, \aap, 631, A153, \dodoi{10.1051/0004-6361/201936635}

\bibitem[{{de Wijn} {et~al.}(2022){de Wijn}, {Casini}, {Carlile}, {Lecinski}, {Sewell}, {Zmarzly}, {Eigenbrot}, {Beck}, {W{\"o}ger}, \& {Kn{\"o}lker}}]{2022SoPh..297...22D}
{de Wijn}, A.~G., {Casini}, R., {Carlile}, A., {et~al.} 2022, \solphys, 297, 22, \dodoi{10.1007/s11207-022-01954-1}

\bibitem[{{del Toro Iniesta}(2007)}]{2007insp.book.....D}
{del Toro Iniesta}, J.~C. 2007, {Introduction to Spectropolarimetry}

\bibitem[{{Fontenla} {et~al.}(1993){Fontenla}, {Avrett}, \& {Loeser}}]{1993ApJ...406..319F}
{Fontenla}, J.~M., {Avrett}, E.~H., \& {Loeser}, R. 1993, \apj, 406, 319, \dodoi{10.1086/172443}

\bibitem[{{Gudiksen} {et~al.}(2011){Gudiksen}, {Carlsson}, {Hansteen}, {Hayek}, {Leenaarts}, \& {Mart{\'\i}nez-Sykora}}]{2011A&A...531A.154G}
{Gudiksen}, B.~V., {Carlsson}, M., {Hansteen}, V.~H., {et~al.} 2011, \aap, 531, A154, \dodoi{10.1051/0004-6361/201116520}

\bibitem[{{Hanaoka}(2005)}]{2005PASJ...57..235H}
{Hanaoka}, Y. 2005, \pasj, 57, 235, \dodoi{10.1093/pasj/57.1.235}

\bibitem[{Harris {et~al.}(2020)Harris, Millman, van~der Walt, Gommers, Virtanen, Cournapeau, Wieser, Taylor, Berg, Smith, Kern, Picus, Hoyer, van Kerkwijk, Brett, Haldane, del R{\'{i}}o, Wiebe, Peterson, G{\'{e}}rard-Marchant, Sheppard, Reddy, Weckesser, Abbasi, Gohlke, \& Oliphant}]{harris2020array}
Harris, C.~R., Millman, K.~J., van~der Walt, S.~J., {et~al.} 2020, Nature, 585, 357, \dodoi{10.1038/s41586-020-2649-2}

\bibitem[{Hunter(2007)}]{Hunter2007}
Hunter, J.~D. 2007, Computing in Science \& Engineering, 9, 90, \dodoi{10.1109/MCSE.2007.55}

\bibitem[{{Jaume Bestard} {et~al.}(2022){Jaume Bestard}, {Trujillo Bueno}, {Bianda}, {{\v{S}}t{\v{e}}p{\'a}n}, \& {Ramelli}}]{2022A&A...659A.179J}
{Jaume Bestard}, J., {Trujillo Bueno}, J., {Bianda}, M., {{\v{S}}t{\v{e}}p{\'a}n}, J., \& {Ramelli}, R. 2022, \aap, 659, A179, \dodoi{10.1051/0004-6361/202141834}

\bibitem[{{Joshi} \& {Rouppe van der Voort}(2022)}]{2022A&A...664A..72J}
{Joshi}, J., \& {Rouppe van der Voort}, L. H.~M. 2022, \aap, 664, A72, \dodoi{10.1051/0004-6361/202243051}

\bibitem[{{Kurucz}(2011)}]{2011CaJPh..89..417K}
{Kurucz}, R.~L. 2011, Canadian Journal of Physics, 89, 417, \dodoi{10.1139/p10-104}

\bibitem[{{Lagg} {et~al.}(2017){Lagg}, {Lites}, {Harvey}, {Gosain}, \& {Centeno}}]{2017SSRv..210...37L}
{Lagg}, A., {Lites}, B., {Harvey}, J., {Gosain}, S., \& {Centeno}, R. 2017, \ssr, 210, 37, \dodoi{10.1007/s11214-015-0219-y}

\bibitem[{{Landi Degl'Innocenti} \& {Landolfi}(2004)}]{2004ASSL..307.....L}
{Landi Degl'Innocenti}, E., \& {Landolfi}, M. 2004, {Polarization in Spectral Lines}, Vol. 307, \dodoi{10.1007/978-1-4020-2415-3}

\bibitem[{{Leenaarts} \& {Carlsson}(2009)}]{2009ASPC..415...87L}
{Leenaarts}, J., \& {Carlsson}, M. 2009, in Astronomical Society of the Pacific Conference Series, Vol. 415, The Second Hinode Science Meeting: Beyond Discovery-Toward Understanding, ed. B.~{Lites}, M.~{Cheung}, T.~{Magara}, J.~{Mariska}, \& K.~{Reeves}, 87

\bibitem[{{Leenaarts} {et~al.}(2012){Leenaarts}, {Carlsson}, \& {Rouppe van der Voort}}]{2012ApJ...749..136L}
{Leenaarts}, J., {Carlsson}, M., \& {Rouppe van der Voort}, L. 2012, \apj, 749, 136, \dodoi{10.1088/0004-637X/749/2/136}

\bibitem[{{L{\'o}pez Ariste} {et~al.}(2005){L{\'o}pez Ariste}, {Casini}, {Paletou}, {Tomczyk}, {Lites}, {Semel}, {Landi Degl'Innocenti}, {Trujillo Bueno}, \& {Balasubramaniam}}]{2005ApJ...621L.145L}
{L{\'o}pez Ariste}, A., {Casini}, R., {Paletou}, F., {et~al.} 2005, \apjl, 621, L145, \dodoi{10.1086/429158}

\bibitem[{{Mathur} {et~al.}(2022){Mathur}, {Joshi}, {Nagaraju}, {Rouppe van der Voort}, \& {Bose}}]{2022A&A...668A.153M}
{Mathur}, H., {Joshi}, J., {Nagaraju}, K., {Rouppe van der Voort}, L., \& {Bose}, S. 2022, \aap, 668, A153, \dodoi{10.1051/0004-6361/202244332}

\bibitem[{{Mathur} {et~al.}(2023){Mathur}, {Nagaraju}, {Joshi}, \& {de la Cruz Rodr{\'\i}guez}}]{2023ApJ...946...38M}
{Mathur}, H., {Nagaraju}, K., {Joshi}, J., \& {de la Cruz Rodr{\'\i}guez}, J. 2023, \apj, 946, 38, \dodoi{10.3847/1538-4357/acbf49}

\bibitem[{{Mathur} {et~al.}(2024){Mathur}, {Nagaraju}, {Yadav}, \& {Joshi}}]{2024ApJ...971...30M}
{Mathur}, H., {Nagaraju}, K., {Yadav}, R., \& {Joshi}, J. 2024, \apj, 971, 30, \dodoi{10.3847/1538-4357/ad54ba}

\bibitem[{{Morosin} {et~al.}(2020){Morosin}, {de la Cruz Rodr{\'\i}guez}, {Vissers}, \& {Yadav}}]{2020A&A...642A.210M}
{Morosin}, R., {de la Cruz Rodr{\'\i}guez}, J., {Vissers}, G. J.~M., \& {Yadav}, R. 2020, \aap, 642, A210, \dodoi{10.1051/0004-6361/202038754}

\bibitem[{{Nagaraju} {et~al.}(2008){Nagaraju}, {Sankarasubramanian}, \& {Rangarajan}}]{2008ApJ...678..531N}
{Nagaraju}, K., {Sankarasubramanian}, K., \& {Rangarajan}, K.~E. 2008, \apj, 678, 531, \dodoi{10.1086/533433}

\bibitem[{{Nagaraju} {et~al.}(2020){Nagaraju}, {Sankarasubramanian}, \& {Rangarajan}}]{2020JApA...41...10N}
---. 2020, Journal of Astrophysics and Astronomy, 41, 10, \dodoi{10.1007/s12036-020-9627-9}

\bibitem[{{Pastor Yabar} {et~al.}(2020){Pastor Yabar}, {Mart{\'\i}nez Gonz{\'a}lez}, \& {Collados}}]{2020A&A...635A.210P}
{Pastor Yabar}, A., {Mart{\'\i}nez Gonz{\'a}lez}, M.~J., \& {Collados}, M. 2020, \aap, 635, A210, \dodoi{10.1051/0004-6361/202037480}

\bibitem[{{Przybilla} \& {Butler}(2004)}]{2004ApJ...609.1181P}
{Przybilla}, N., \& {Butler}, K. 2004, \apj, 609, 1181, \dodoi{10.1086/421316}

\bibitem[{{Rempel}(2012)}]{2012ApJ...750...62R}
{Rempel}, M. 2012, \apj, 750, 62, \dodoi{10.1088/0004-637X/750/1/62}

\bibitem[{{Rimmele} {et~al.}(2020){Rimmele}, {Warner}, {Keil}, {Goode}, {Kn{\"o}lker}, {Kuhn}, {Rosner}, {McMullin}, {Casini}, {Lin}, {W{\"o}ger}, {von der L{\"u}he}, {Tritschler}, {Davey}, {de Wijn}, {Elmore}, {Fehlmann}, {Harrington}, {Jaeggli}, {Rast}, {Schad}, {Schmidt}, {Mathioudakis}, {Mickey}, {Anan}, {Beck}, {Marshall}, {Jeffers}, {Oschmann}, {Beard}, {Berst}, {Cowan}, {Craig}, {Cross}, {Cummings}, {Donnelly}, {de Vanssay}, {Eigenbrot}, {Ferayorni}, {Foster}, {Galapon}, {Gedrites}, {Gonzales}, {Goodrich}, {Gregory}, {Guzman}, {Guzzo}, {Hegwer}, {Hubbard}, {Hubbard}, {Johansson}, {Johnson}, {Liang}, {Liang}, {McQuillen}, {Mayer}, {Newman}, {Onodera}, {Phelps}, {Puentes}, {Richards}, {Rimmele}, {Sekulic}, {Shimko}, {Simison}, {Smith}, {Starman}, {Sueoka}, {Summers}, {Szabo}, {Szabo}, {Wampler}, {Williams}, \& {White}}]{2020SoPh..295..172R}
{Rimmele}, T.~R., {Warner}, M., {Keil}, S.~L., {et~al.} 2020, \solphys, 295, 172, \dodoi{10.1007/s11207-020-01736-7}

\bibitem[{{Rutten}(1988)}]{1988ASSL..138..185R}
{Rutten}, R.~J. 1988, in Astrophysics and Space Science Library, Vol. 138, IAU Colloq. 94: Physics of Formation of FE II Lines Outside LTE, ed. R.~{Viotti}, A.~{Vittone}, \& M.~{Friedjung}, 185--210, \dodoi{10.1007/978-94-009-4023-9_23}

\bibitem[{{Rutten} \& {Kostik}(1982)}]{1982A&A...115..104R}
{Rutten}, R.~J., \& {Kostik}, R.~I. 1982, \aap, 115, 104

\bibitem[{{Smitha} {et~al.}(2023){Smitha}, {van Noort}, {Solanki}, \& {Castellanos Dur{\'a}n}}]{2023A&A...669A.144S}
{Smitha}, H.~N., {van Noort}, M., {Solanki}, S.~K., \& {Castellanos Dur{\'a}n}, J.~S. 2023, \aap, 669, A144, \dodoi{10.1051/0004-6361/202245130}

\bibitem[{{Socas-Navarro} \& {Uitenbroek}(2004)}]{2004ApJ...603L.129S}
{Socas-Navarro}, H., \& {Uitenbroek}, H. 2004, \apjl, 603, L129, \dodoi{10.1086/383147}

\bibitem[{{Uitenbroek}(2001)}]{2001ApJ...557..389U}
{Uitenbroek}, H. 2001, \apj, 557, 389, \dodoi{10.1086/321659}

\bibitem[{{V{\"o}gler} {et~al.}(2005){V{\"o}gler}, {Shelyag}, {Sch{\"u}ssler}, {Cattaneo}, {Emonet}, \& {Linde}}]{2005A&A...429..335V}
{V{\"o}gler}, A., {Shelyag}, S., {Sch{\"u}ssler}, M., {et~al.} 2005, \aap, 429, 335, \dodoi{10.1051/0004-6361:20041507}

\bibitem[{{{\v{S}}t{\v{e}}p{\'a}n} \& {Trujillo Bueno}(2010)}]{2010MmSAI..81..810S}
{{\v{S}}t{\v{e}}p{\'a}n}, J., \& {Trujillo Bueno}, J. 2010, \memsai, 81, 810.
\newblock \doarXiv{1001.2720}

\bibitem[{{{\v{S}}t{\v{e}}p{\'a}n} \& {Trujillo Bueno}(2011)}]{2011ApJ...732...80S}
---. 2011, \apj, 732, 80, \dodoi{10.1088/0004-637X/732/2/80}

\bibitem[{{Yadav} {et~al.}(2021){Yadav}, {D{\'\i}az Baso}, {de la Cruz Rodr{\'\i}guez}, {Calvo}, \& {Morosin}}]{2021A&A...649A.106Y}
{Yadav}, R., {D{\'\i}az Baso}, C.~J., {de la Cruz Rodr{\'\i}guez}, J., {Calvo}, F., \& {Morosin}, R. 2021, \aap, 649, A106, \dodoi{10.1051/0004-6361/202039857}

\end{thebibliography}
\bibliographystyle{aasjournal}

\appendix

\section{Detailed analysis of response functions} \label{sect:appresp}

\begin{figure*}[htbp]
    \centering
    \includegraphics[width=\textwidth]{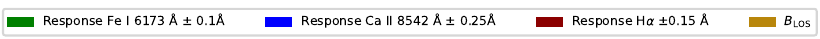}
    \includegraphics[width=\textwidth]{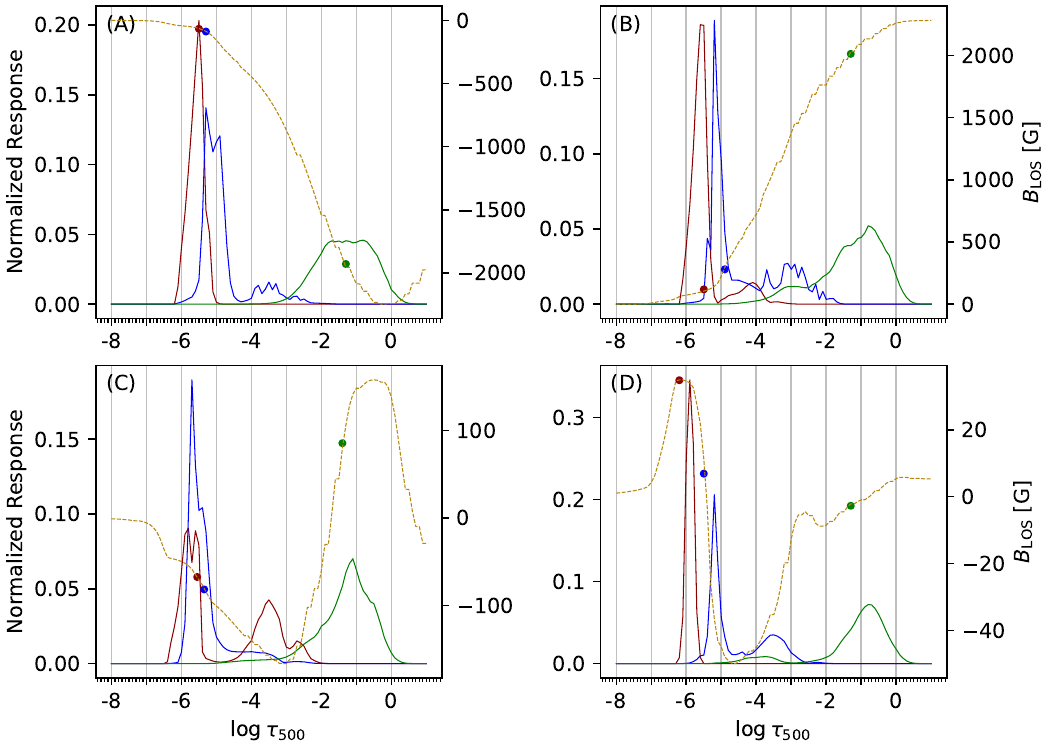}
    \caption{Wavelength integrated normalized response functions of the Stokes~$V$ profiles to the perturbations of the LOS magnetic field for the \FeIline{}, \CaIR{} and \Halpha{} lines. The response functions for the profiles for the 4-selected pixels shown in Fig.~\ref{fig:profcontext} are shown in panels (A), (B), (C) and (D), respectively. The stratification of the \blos{} is also shown for completeness. The value of the LOS magnetic field inferred from the \Halpha{}, \CaIR{} and \FeIline{} spectral lines is overplotted with $\bullet$ at \logtau{} values with closest match with the rMHD model, in the same color scheme as the spectral lines.}
    \label{fig:response}
\end{figure*}

To determine the optical depths (heights) at which the Stokes~$V$ profile of the \Halpha{} line is most sensitive to the magnetic field, we have analyzed the response of the Stokes~$V$ profiles to the perturbations of the LOS magnetic field.
The details about the calculation of the response functions are mentioned in section~\ref{sect:resp}.
In Fig.~\ref{fig:response}, we show wavelength-integrated response functions of the Stokes~$V$ profiles to the perturbations of the LOS magnetic field for the \FeIline{} (green-colored curve) line, \CaIR{} (blue-colored curve) line, and the \Halpha{} (red-colored curve) line, for the 4-selected pixels marked in Fig.~\ref{fig:context}.
The integration wavelength range is set to $\pm$0.1~\AA\, for the \FeIline{} line, $\pm$0.25~\AA\, for the \CaIR{} line, and $\pm$0.15~\AA\, for the \Halpha{} line.
For comparison, the stratification of the LOS magnetic field from the 3D MHD model is overplotted (brown-colored curve) and the scale is marked on the right-axis.
The magnetic field values inferred from the synthetic spectral profiles of the \Halpha{}, \CaIR{} and \FeIline{} lines with $\bullet$ at \logtau{} values with the closest match with the rMHD model are also indicated.

In panels (A) and (B), we show the response functions of the Stokes~$V$ profiles to the perturbations of the LOS magnetic field for pixels where the magnetic field is consistently positive (or negative) polarity across all atmospheric heights.
The \Halpha{} line exhibits its peak sensitivity to LOS magnetic field perturbations at \logtau{} = $-$5.6, while the \CaIR{} line reaches its maximum response at \logtau{} = $-$5.1.
For the \FeIline{} line, the peak response occurs at \logtau{} = $-$1.2.
The LOS magnetic field inferred from the \Halpha{}, \CaIR{}, and \FeIline{} lines closely match with the value of the rMHD model at the \logtau{} values at which their respective response functions are maximum.

The atmospheres of the remaining two selected pixels exhibit complex LOS magnetic field stratification, characterized by steep gradients and polarity changes with height, as shown in panels (C) and (D).
In the atmosphere of the pixel shown in panel (C), the LOS magnetic field is positive in the lower photospheric layers (\logtau{} $>$ $-$1.8) and negative in the upper photospheric and chromospheric layers (\logtau{} $<$ $-$1.8).
Whereas, in the atmosphere of the pixel shown in panel (D), the LOS magnetic field is negative in the photosphere and lower chromosphere (\logtau{} $>$ $-$5.4) and positive in the upper chromosphere (\logtau{} $<$ $-$5.4).

In the case shown in panel (C), where the LOS magnetic field exhibits opposite polarities in the lower photosphere and chromosphere, the \Halpha{} and \CaIR{} lines exhibit peak responses to LOS magnetic field perturbations around \logtau{} = $-$5.8 and $-$5.7, respectively.
Additionally, the \Halpha{} line displays a notable response around \logtau{} = $-$3.5.
The \FeIline{} line peaks around \logtau{} = $-$1.2.
The inferred LOS magnetic fields from the \Halpha{}, \CaIR{} and \FeIline{} lines closely match with the rMHD model, however, they differ from the values of the rMHD model at the \logtau{} where the corresponding response is maximum.

For the case shown in panel (D), where the LOS magnetic field has opposite polarities in the lower and upper chromosphere, the \Halpha{} line exhibits peak response to LOS magnetic field perturbations at \logtau{} = $-$5.9.
While the \CaIR{} and the \FeIline{} lines peak at \logtau{} = $-$5.1 and $-$0.8, respectively.
Similar to the scenario depicted in panel (C), the LOS magnetic field values inferred from the \Halpha{}, \CaIR{}, and \FeIline{} lines are different compared to the value of the rMHD model at the \logtau{} where their respective response functions are maximum.

As discussed in section~\ref{sect:mapping} and shown in Fig.~\ref{fig:tauformation}, the LOS magnetic field map inferred from the \Halpha{} line core closely corresponds to the rMHD model at \logtau{} = $-$5.7.
However, the analysis of the response functions suggests that, for individual pixels, the maximum response of the Stokes~$V$ profile to LOS magnetic field perturbations can occur at different \logtau{} values, depending on the magnetic field stratification.

For relatively simple stratifications, where the LOS magnetic field maintains the same polarity with height, the inferred magnetic field closely matches the value of the rMHD model at the \logtau{} where the response is maximum.
In contrast, for pixels with complex stratification—characterized by steep gradients and polarity reversals—the inferred LOS magnetic field remains close to the rMHD model but may be different than the value of the rMHD model at the \logtau{} with the maximum response.
This discrepancy may result from the complex Stokes~$I$ and $V$ profiles observed for these pixels, see brown and red-colored \si{} and \sv{} profiles of the \CaIR{} line of Fig.~\ref{fig:profcontext}.
In addition to the complex LOS magnetic field stratification, the temperature structure and the velocity fields may have contributed to the formation of such profiles.
Interestingly, for the pixel in panel (D), restricting the WFA analysis of the \CaIR{} line to only the blue wing range ([$-$0.25, 0]~\AA) yields a LOS magnetic field of approximately $-$20~G, which corresponds to the value of the rMHD model at the \logtau{} of maximum response ($-$5.1).

Overall, the response function analysis clearly demonstrates that the maximum response of the Stokes~$V$ profiles of the \Halpha{} line to the perturbations of the LOS magnetic field occurs at higher atmospheric layers than that of the \CaIR{} line.
Consequently, the LOS magnetic fields inferred from the Stokes~$V$ profiles of the \Halpha{} line core are from higher atmospheric layers compared to those inferred from \CaIR{} line.

\section{Validation of WFA under simulated noise conditions} \label{sect:dkist}

\begin{figure*}[htbp]
    \centering
    \includegraphics[]{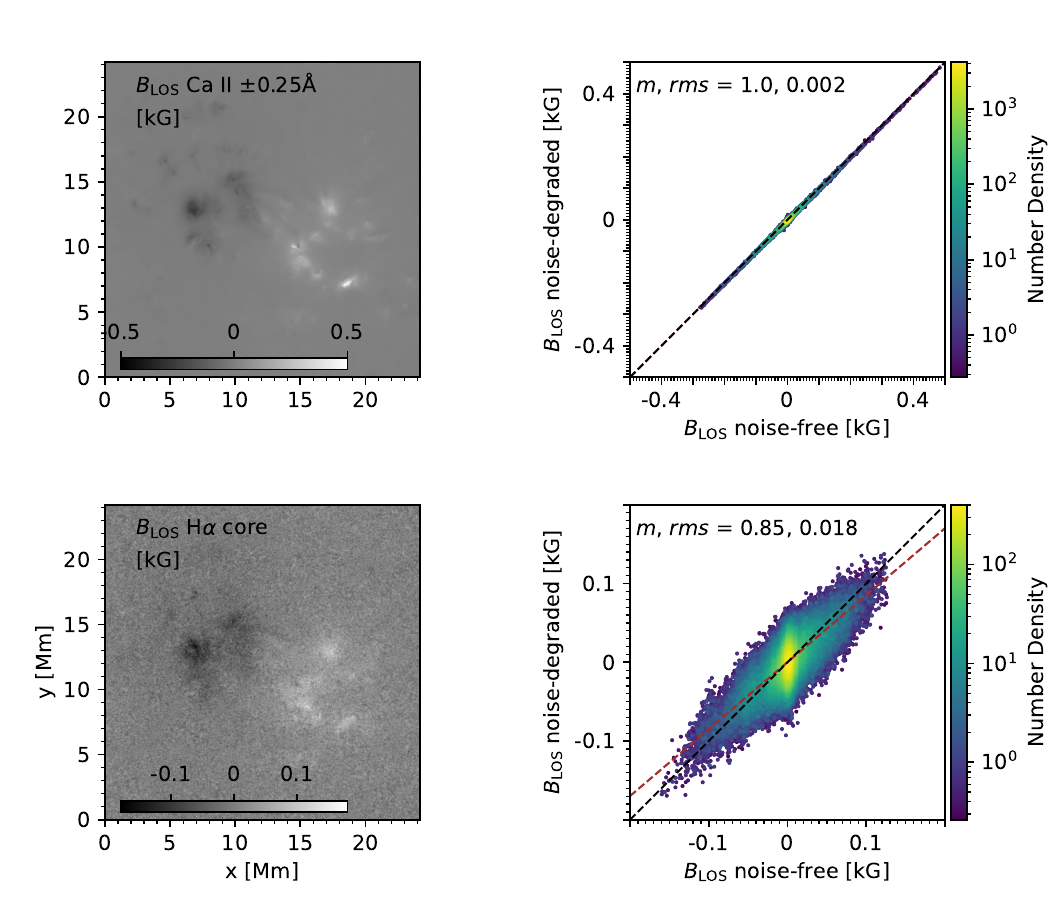}
    \caption{Top-left: Map of the line-of-sight magnetic field inferred from the \CaIR{} line using the WFA applied to synthetic Stokes profiles degraded with Gaussian noise ($\sigma$ = 10$^{-3} I_c$). Top-right: Scatter plot comparing WFA-inferred \blos{} from noise-free versus noise-degraded \CaIR{} profiles. Bottom panels: Same as top, but for the \Halpha{} line. The slope of the linear regression of the scatter plot in the third column is indicated by $m$ and rms indicates the root mean square of the difference between the two values.}
    \label{fig:dkist}
\end{figure*}

To estimate the effect of noise in high-resolution ground-based observations, we introduce Gaussian noise with a standard deviation of $\sigma = 10^{-3}I_c$ into the synthetic Stokes profiles and re-infer the LOS magnetic field maps using the WFA, following the same procedure as in Fig.~\ref{fig:fecacomparisonmap} and \ref{fig:hacomparisonmap}.
Figure~\ref{fig:dkist} displays the resulting inferred maps, accompanied by scatter plots comparing the fields retrieved from the noise-free and noise-degraded profiles.
For the \CaIR{} line, the magnetic field maps derived from the noise-degraded profiles show minimal deviation from those obtained in the noise-free case.
This robustness is likely due to the relatively stronger \sv{} signals in the \CaIR{} line, which reach amplitudes of approximately 10\%.
In contrast, the \Halpha{} maps exhibit significantly more noise when inferred from noise-degraded profiles.
We attribute this to the inherently weaker \sv{} amplitudes in \Halpha{} line, which are on the order of $\sim$0.2\%.
Despite the increased noise, the underlying magnetic structures remain discernible and coherent.
The rms error between the noise-degraded and noise-free maps in the \Halpha{} case is found to be 18~G.
Notably, the scatter decreases with increasing field strength.
It is important to highlight that the noisiness of the \Halpha{} magnetic field map primarily arises from the properties of the underlying simulation, which models an enhanced network quiet-Sun atmosphere with maximum field strengths of only $\sim$110~G at \logtau{} = $-$5.7.
In observational scenarios involving stronger magnetic fields, such as active regions studied by \citet{2023ApJ...946...38M, 2024ApJ...971...30M}, where chromospheric field strengths can reach 500--1000~G, the polarization signals would be substantially stronger and more reliably detected.

\end{document}